\documentclass[conference]{IEEEtran}
\ifCLASSINFOpdf
  \usepackage[pdftex]{graphicx}
\else
\fi
\usepackage{amsmath}
\usepackage{url}

\usepackage[final]{changes}
\usepackage{algpseudocode}
\usepackage{algorithm}
\usepackage{todonotes}  
\usepackage{soul}  

\newif\ifdraftmode
\draftmodefalse 

\begin{document}
%
\title{ \replaced{Facilitating AI and System Operator Synergy: Active Learning-Enhanced Digital Twin Architecture for Day-Ahead Load Forecasting}{Developing an AI-Assistant for System Operators: Active Learning-Enhanced Digital Twin for Optimized Power Grid Management} }

\author{\IEEEauthorblockN{Costas Mylonas, Titos Georgoulakis, Magda Foti}
\IEEEauthorblockA{\textit{UBITECH}, 
Thessalias 8,
15231 Chalandri,
Athens, Greece\\
\{kmylonas, tgeorgoulakis, mfoti\}@ubitech.eu}
}


%


\maketitle

\begin{abstract}
In this paper, we introduce a synergistic approach between artificial intelligence and system operators through an innovative digital twin architecture, integrated with an active learning framework, to enhance short-term load forecasting. Central to this architecture is the incorporation of sophisticated data pipelines, facilitating the real-time ingestion, processing and analysis of grid-related data. Utilizing a recurrent neural network architecture, our model generates day-ahead load forecasts together with prediction \added{confidence} intervals, strengthening system operator trust in the model's predictive reliability and enhancing their ability to respond to evolving grid conditions effectively. The active learning framework iteratively refines the predictions by incorporating real-time feedback based on forecast uncertainty, \replaced{utilizing}{harnessing} newly available data to continuously enhance forecasting accuracy and confidence. This AI-assisted strategy is exemplified in a case study of the Greek transmission system. It demonstrates the potential to transform short-term load forecasting, thereby increasing the reliability and operational efficiency of modern power grids. This approach marks a significant step forward in the digitalization and intelligent management of power systems.
\end{abstract}


%
\IEEEpeerreviewmaketitle

\section{Introduction} \label{sec:Intro}

The rapid evolution of power systems, driven by the rapid digitalization and the shift towards renewable energy, poses significant challenges in grid management. This transformation necessitates the integration of more intelligent and responsive frameworks to ensure stable and efficient grid operations. The role of Artificial Intelligence (AI) in assisting grid operators has become increasingly crucial in navigating these complexities \added{\cite{chen2024artificial}}. AI's potential to enhance decision-making under uncertainty and its application in grid management is a growing area of interest.

Highlighting the complexities of power grid operations in the digital age, 
\replaced{recent studies stress the need for advanced AI-driven Human-Machine Interfaces (HMI) \cite{marot:hal-03123117}, promoting new frameworks for grid management assistance and discuss the changing roles of human operators in control rooms \cite{mazumder2024power}, emphasizing the cognitive challenges and decision support systems needed in highly automated power systems.}{\cite{marot:hal-03123117} stresses the need for advanced AI-driven Human-Machine Interfaces (HMI), promoting new frameworks for grid management assistance. \cite{prostejovsky2019future} discusses the changing roles of human operators in control rooms, emphasizing the cognitive challenges and decision support systems needed in highly automated power systems.}

Innovations in Digital Twin (DT) technology are revolutionizing smart grids. Study \cite{deakin2023smart} introduces a DT framework for electrical distribution systems, emphasizing its practical application in addressing the integration of imperfect data, which is crucial for realistic DT implementations in distribution networks. Complementing this, \cite{zhou2019digital} details a DT framework for power grid online analysis, focusing on its integration with an actual power grid's Energy Management System (EMS) and highlighting features like in-memory computing and machine learning. This framework demonstrates the potential of DTs in enhancing decision-making and operational efficiency in power grid management. Together, these studies highlight the transformative role of DT technology in advancing smart grid operations.

Probabilistic Load Forecasting (PLF) has become a vital tool for electricity market participants and system operators, particularly for anticipating grid challenges like power imbalances and congestions. \cite{gurses2020probabilistic} \replaced{explores}{delves into} this field with a Recurrent Neural Network (RNN) designed for day-ahead forecasting of residual loads. Their approach includes both parametric and non-parametric models, ensuring reliable forecasts. Key to their methodology is the use of probabilistic evaluation metrics like the ignorance score and quantile score, enhancing the model's accuracy and facilitating its comparison with other forecasting methods. Complementing this, \cite{browell2020probcast} introduces ProbCast, a versatile tool for generating probabilistic forecasts, especially in energy forecasting. ProbCast supports advanced techniques like parametric and non-parametric density forecasting, making it instrumental in managing uncertainties in power system operations. 

In the smart grid domain, Active Learning (AL) is enhancing the adaptability and the accuracy of forecasting models. \cite{wang2019deep} developed a deep ensemble learning model for short-term load forecasting (STLF), which employs an AL framework. This model integrates a Long Short-Term Memory (LSTM) network with a multi-layer perceptron to accurately capture the \replaced{complex}{intricate} load patterns affected by various factors like weather. The AL component selectively trains the model using similar load segments, effectively addressing data imbalances and enhancing forecasting performance. \cite{zhang2021active} explored an AL strategy for building energy forecasting, efficiently generating informative training data while considering weather impacts. Their approach successfully addresses data bias problems common in building operation data, leading to improved model accuracy and extendibility, thus showcasing the potential of AL in energy management and forecasting.

This paper introduces an innovative approach that facilitates the synergy between AI and system operators through a novel DT architecture integrated with an AL framework for enhanced STLF. \added{Our approach not only aligns with the trajectory of the papers reviewed but also extends their individual contributions into a comprehensive solution for solving modern grid challenges, distinct in its integration of these concepts into a cohesive system that advances beyond the individual solutions presented in the literature.} \deleted{It does not only align with the trajectory of the papers reviewed, but also extends their individual contributions into a comprehensive, dynamic solution for solving modern grid challenges. Our approach is distinct in its integration of these concepts into a cohesive system, advancing beyond the individual solutions presented in the literature.} \added{While this paper focuses on the integration of PLF as a key service, the proposed DT architecture is designed to support various AI-driven services, thus offering a versatile platform for intelligent grid management.}

The rest of this paper is structured as follows: Section \ref{sec:DTArchitecture} elaborates on the DT architecture, Section \ref{sec:PLF} discusses the PLF, Section \ref{sec:ALFramework} \replaced{explains}{delves into} the AL framework, Section \ref{sec:Results} presents a detailed case study with results, and Section \ref{sec:Conlusion} concludes with a summary of our findings and future research directions.

\section{Digital Twin Architecture} \label{sec:DTArchitecture}

\begin{figure}[h]
    \centering
    \includegraphics[width=\columnwidth]{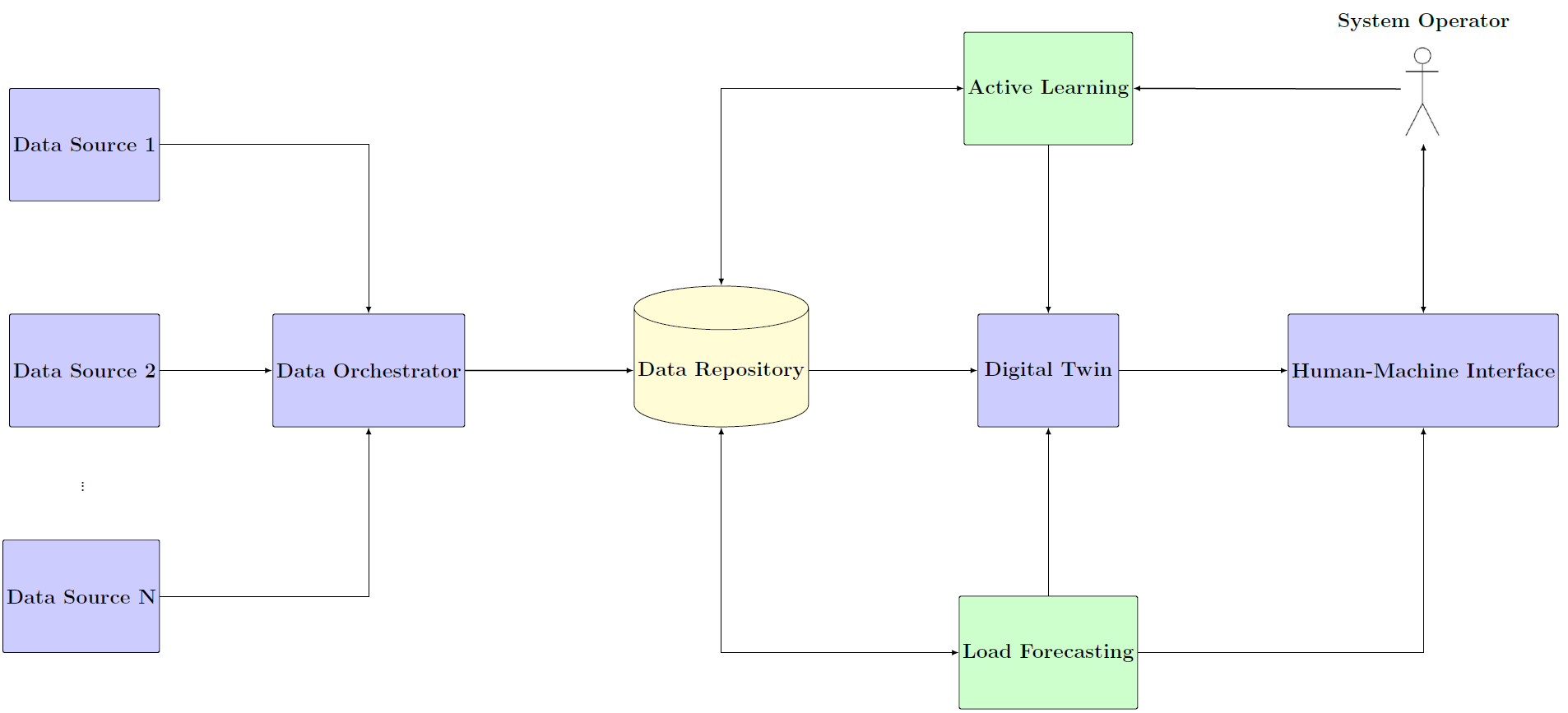}
    \caption{DT Architecture.} 
    \label{fig:DTArchitecture}
\end{figure}

The DT architecture presented in this paper is a combination of data management and computational modeling and simulation of power grid networks to improve the operational decision-making processes. \added{It brings together real-time and historical data with advanced computational analytics, providing a robust digital replica of the power grid for enhanced management and decision-making.} As illustrated in Figure \ref{fig:DTArchitecture}, the architecture consists of several key components, each serving a distinct role within the system.

Data sources act as inputs, varying from CSV files to real-time data streams provided by APIs. These sources encompass various time-series data, including day-ahead load and generation forecasts, actual load and generation, weather-related data, and static data, such as grid topologies and infrastructure specifications. Dagster \cite{dagster}, a modern framework that orchestrates the flow of data from these disparate sources, serves as the cornerstone for data ingestion, pre-processing and storage. Dagster's primary role is to streamline the data lifecycle processes, namely extraction, transformation, and loading to facilitate timely and organized data delivery to the TimescaleDB \cite{timescaledb}. This time-series database is suitable for managing large-scale data with intrinsic temporal attributes, ensuring data fidelity and query efficiency.

At the core of the DT architecture is PyPSA \cite{PyPSA}, a comprehensive tool for power system analysis that enables network modeling and power flow solving. PyPSA is the computational engine that enables the simulation of power grid behavior under different operational scenarios.

OperatorFabric \cite{operatorfabric}, a state-of-the-art HMI that presents system operators with intuitive access to real-time insights and simulation outputs, acts as the visualization engine. This interface is crucial for the operators to make informed decisions based on the simulations and analytics results provided by PyPSA.

\added{The DT architecture is designed to support various services on top of its core structure. These services, in our case Load Forecasting and Active Learning, interact with the data repository for data storage and retrieval, providing their outputs to the DT for scenario simulation and to the HMI for visualization. The system operator supervises the results through the HMI and sets parameters for the AL based on their experience, establishing a human-in-the-loop approach.}

\deleted{The DT architecture forms a cohesive and integrated methodology that brings together real-time and historical data with advanced computational analytics, providing a robust platform for enhanced grid management and operational decision-making.}

\section{Probabilistic Load Forecasting} \label{sec:PLF}

The basis of our approach towards AI-assisted decision-making to enhance grid management is the implementation of PLF utilizing RNNs. RNNs, particularly effective in forecasting time-series data due to their ability to handle variable-length sequences and share weights across time steps, form the backbone of our probabilistic model.

\subsection{Model Architecture\deleted{, Inputs and Loss Function}}

\begin{figure}[h]
    \centering
    \includegraphics[width=\columnwidth]{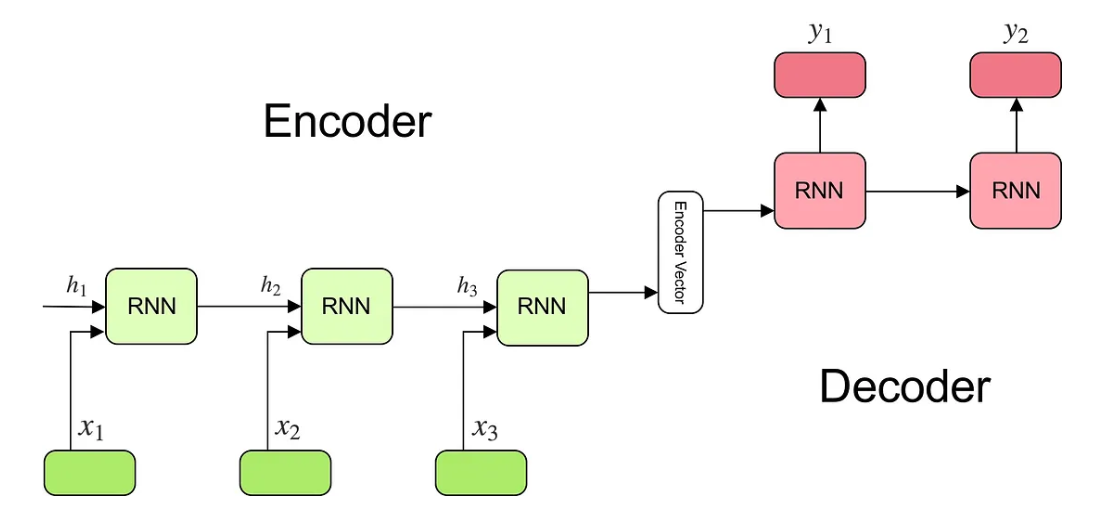}
    \caption{RNN Architecture: Encoder-Decoder Implementation.}
    \label{fig:RNNarchitecture}
\end{figure}

Our PLF model leverages an advanced RNN architecture \cite{sherstinsky2020fundamentals}, specifically designed to handle the intricacies of load forecasting with a particular emphasis on the probabilistic part enforced through a loss function that takes into account both forecast accuracy and uncertainty. By adopting an encoder-decoder implementation, \added{as can be seen in Figure \ref{fig:RNNarchitecture},} our approach processes sequentially time-series data, capturing temporal dependencies essential for accurate forecasting.

The encoder sequentially processes past input data \added{(\(x_1, x_2, x_3, \ldots\))}, such as historical load profiles and various meteorological conditions, extracting important patterns. \added{Each input \(x_t\) at time step \(t\) generates a hidden state \(h_t\) through the RNN cell.} This information is encapsulated within the RNN cell state constituting a compact internal representation of historical data insights. \added{ The final hidden state from the encoder, known as the encoder vector, summarizes all the input information up to the current time step.} Subsequently, the decoder, informed by the encoder's state, combines this knowledge with additional inputs \added{(\(y_1, y_2, \ldots\))} to predict future loads. These supplementary inputs include time-based variables, such as the hour of the day and the month, providing to the model precise temporal context. The decoder's predictive performance is enhanced by a fully-connected neural network incorporating dropout, which adds robustness to the forecasting task. The features are normalized using MinMax scaling to ensure that the model inputs have a consistent scale, which can aid in the convergence and performance of the model.

Targeting accurate load forecasts and confidence estimation, our model employs a Gaussian Negative Log Likelihood (GNLL) loss function, defined as:

\begin{equation}
    GNLL = \frac{1}{T} \sum_{t=1}^{T} \left[ \frac{1}{2} \ln(2\pi\sigma_t^2) + \frac{(y_t - \mu_t)^2}{2\sigma_t^2} \right],
\end{equation}

\noindent where $T$ is the number of time steps, $\mu_t$ and $\sigma_t^2$ represent the forecasted mean and variance respectively, and $y_t$ denotes the true load at time $t$. This loss function rigorously quantifies the model's performance in capturing the distribution of load forecasts, facilitating the generation of reliable and accurate prediction intervals.

Moreover, the RNN model's implementation in PyTorch \cite{paszke2019pytorch} leverages dynamic computational graphs, allowing for flexible coding and efficient training via GPU acceleration. This adaptability is essential for iterative model refinement contributing to the development of a robust PLF model capable of addressing the challenges posed by the evolving energy landscape.

\subsection{Probabilistic Forecast Metrics}

\added{Probabilistic forecasting extends beyond simple point predictions by providing a comprehensive statistical distribution of future events, characterized by Probability Density Functions (PDF) or Cumulative Distribution Functions (CDF), allowing system operators to evaluate and manage risks. These methods can be categorized into non-parametric and parametric approaches. Non-parametric methods derive a set of quantile values by minimizing quantile/pinball loss without assuming a predefined distribution shape. Parametric methods, on the other hand, assume a specific distribution form, such as normal or log-normal, and optimize its parameters by minimizing losses like negative log-likelihood.}

\deleted{Probabilistic forecasting extends beyond simple point predictions to provide a comprehensive statistical distribution of future events, which are expressed through predictive distributions. These distributions can be characterized by Probability Density Functions (PDF) or Cumulative Distribution Functions (CDF), offering a differentiated view of future uncertainties enabling system operators to evaluate and manage the risks associated with different outcomes.}

\deleted{Probabilistic forecasting methods can be categorized into non-parametric and parametric. Non-parametric methods do not assume a predefined shape for the distribution. Instead, they focus on deriving a set of quantile values by minimizing quantile/pinball loss, which measures the cost of deviations from specified quantiles, providing a detailed summary of the predictive CDF. Parametric methods assume a specific distribution form, such as normal, log-normal, etc. They aim to optimize the parameters of the chosen predictive distribution by minimizing losses, such as the negative log-likelihood.}

The evaluation of probabilistic forecasts involves several key metrics, each providing unique insights into the forecast performance. In our paper, the evaluation of probabilistic forecasts specifically employs Prediction Interval Coverage Probability (PICP) and Sharpness. PICP evaluates the percentage of observations that fall within the predicted intervals, serving as a crucial metric for forecast reliability and ensuring that forecasts accurately represent the uncertainty in predictions:
    \begin{equation}
        PICP = \frac{1}{N} \sum_{i=1}^{N} \mathbf{1}_{\{y_i \in [L_i, U_i]\}},
    \end{equation}
\noindent where $N$ is the dataset size, $y_i$ is the actual value, and $U_i$ and $L_i$ are the upper and lower bounds of the prediction interval. Sharpness measures the concentration of predictive distributions, highlighting the precision of the forecasts independent of their actual accuracy. It is quantified by the average width of the central prediction intervals, with a narrower width indicating more precise forecasts. This metric is crucial as it demonstrates the model's capacity to provide detailed forecasts while maintaining reliability:
    \begin{equation}
        Sharpness = \frac{1}{N} \sum_{i=1}^{N} (U_i - L_i),
    \end{equation}

Incorporating these metrics into our evaluation framework ensures that our probabilistic forecasts are accurate and provide meaningful uncertainty estimates, aligning with the real-world complexities of forecasting in power grids. These metrics serve as the foundation for robust model assessment, guiding both the refinement and deployment of our predictive models.

\section{Active Learning Framework} \label{sec:ALFramework}

\begin{figure}[h]
    \centering
    \includegraphics[width=\columnwidth]{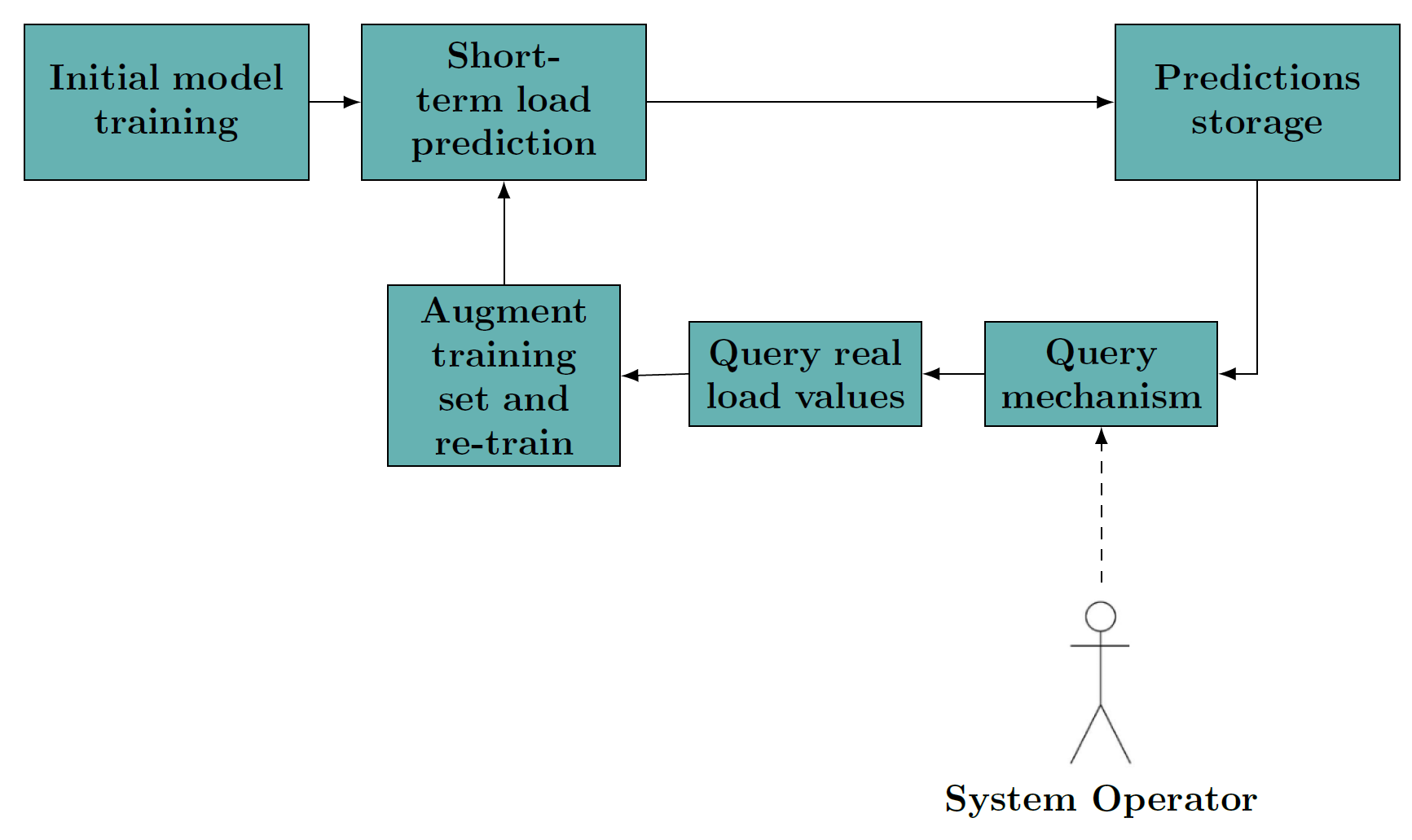}
    \caption{AL Framework.} 
    \label{fig:ALframework}
\end{figure}

The current study introduces an AL framework specifically designed to improve the accuracy and reliability of load forecasting models, \replaced{particularly}{with a particular emphasis on} the RNN model capable of generating confidence intervals, as presented in Section~\ref{sec:PLF}. \deleted{This iterative, dynamic framework significantly improves the model's ability to adapt to the ever-changing landscape of power grid management, where variability is continuous due to factors like renewable energy integration and fluctuating consumption patterns.}\added{This iterative framework enhances the model's adaptability to the dynamic nature of power grid management, influenced by renewable energy integration and fluctuating consumption patterns.} The AL framework includes a sequence of steps, as depicted in Fig.~\ref{fig:ALframework}, beginning with the initial training phase\added{, which establishes a benchmark for subsequent iterative improvements.}\deleted{This foundational stage utilizes a comprehensive dataset, rich in historical load data and weather conditions, establishing a benchmark for subsequent iterative improvements.} \deleted{Following this initial setup,} The model \added{then} \replaced{starts}{embarks on} making short-term load predictions on new data, \deleted{concurrently} quantifying the uncertainty of these predictions \added{and the predictions \(y_t\) together with the mean \(\mu_t\) and the standard deviation \(\sigma_t\) are stored in the data repository}. \deleted{This dual task is pivotal, as it helps identify data points that promise the greatest incremental learning value. By utilizing uncertainty sampling, the framework selects these data points based on the model's exhibited uncertainty. For each forecast \(y_t\), with mean \(\mu_t\) and standard deviation \(\sigma_t\), uncertainty is quantified as \(U_t = \sigma_t\), and high uncertainty instances are those where \(U_t > \theta\), suggesting that learning from these uncertain predictions could significantly enhance the model's forecasting accuracy.}

\added{There are several strategies for the query mechanism in AL, including uncertainty sampling, query by committee, expected model change, expected error reduction, and diversity sampling \cite{tharwat2023survey}. Uncertainty sampling identifies the data points on which the model is least certain. Query by committee uses a committee of models and selects data points where there is the most disagreement among the models. Expected model change selects data points that would cause the most significant change to the current model if labeled. Expected error reduction chooses data points that are expected to reduce the overall error of the model the most. Diversity sampling ensures that the selected data points are diverse and cover different regions of the data space.}

\added{In our case, we use uncertainty sampling and the uncertainty \(U_t\) is quantified through the standard deviation \(\sigma_t\). The query mechanism \(Q\) selects data points for acquisition based on the function:}

\added{
\begin{equation} \label{eqn:4}
Q = \{t | U_t = \sigma_t > \theta\}, 
\end{equation}
}

\added{\noindent where \(\theta\) is set based on empirical analysis by the system operator. The value of \(\theta\) is crucial as it determines the threshold of uncertainty above which data points are queried. Initially, \(\theta\) is set based on historical analysis of the model's performance on past data, and this is often an empirical decision made by the system operator.}

\added{This human-in-the-loop approach ensures that the model's learning process is guided by expert knowledge and operational priorities. The query mechanism identifies data points with high uncertainty and then the framework initiates an automated query process for the actual load values corresponding to these uncertain predictions from the data repository. This step is vital, as it supplies the model with real, observed data that was previously marked by significant predictive uncertainty.} \deleted{The querying mechanism \(Q\) selects data points for acquisition based on the function \(Q = \{t | U_t > \theta\}\), where \(\theta\) is set based on empirical analysis or expert judgment.} With the newly acquired and augmented dataset, the model undergoes a retraining process.\deleted{, allowing it to recalibrate its predictions and adapt to emerging patterns and trends in the dataset. This step integrates the newly acquired information, enabling the model to recalibrate its predictions and adapt to emerging patterns and trends in the dataset.} \added{This process, also described in Algorithm \ref{alg:AL}, allows the model to incorporate the new information, adjust its predictions, and improve its ability to recognize emerging patterns and trends in the data.} \deleted{Through this iterative cycle of prediction, identification, querying, and updating, the AL framework ensures continuous improvement in the model's performance, enhancing its adaptability and responsiveness to the dynamic nature of grid operations.}

\begin{algorithm}
\caption{Active Learning with Uncertainty Sampling}
\label{alg:AL}
\begin{algorithmic}[1]
\State \textbf{Input:} Initial training dataset $\mathcal{D}_{train}$, Uncertainty threshold $\theta$
\State Train initial RNN model on $\mathcal{D}_{train}$
\For{each prediction cycle}
    \State Make predictions $y_t$ with mean $\mu_t$ and standard deviation $\sigma_t$
    \State Store predictions and uncertainties in the data repository
    \State Select high-uncertainty points according to \eqref{eqn:4}
    \State Query actual load values for $Q$ from the data repository
    \State Augment $\mathcal{D}_{train}$ with new data points from $Q$
    \State Retrain RNN model on the augmented $\mathcal{D}_{train}$
    \State Update uncertainty threshold $\theta$ based on new predictions (if necessary)
\EndFor
\end{algorithmic}
\end{algorithm}

\deleted{During the iterative refinement process facilitated by the AL framework, the model's predictions, along with their uncertainties, are critically evaluated. High uncertainty predictions, where \(U_t = \sigma_t\) exceeds the predetermined threshold \(\theta\), are targeted for further data acquisition and model retraining. This approach ensures continuous learning and adaptation, effectively leveraging the GNLL loss function to iteratively enhance both the precision of load forecasts and the reliability of the associated confidence intervals. Through this dynamic interplay between prediction, evaluation, and refinement, the AL framework significantly contributes to the advancement of intelligent grid management practices.}

\added{Through this iterative cycle of predicting, querying, and updating, the AL framework ensures continuous improvement in the model's performance, enhancing its adaptability to the dynamic nature of grid operations. The system operator reviews and adjusts $\theta$ after each cycle, ensuring continuous improvement in the model's performance. Also, the DT architecture gives the ability to the system operators to identify through the HMI data points that correspond to rare events, such as extreme weather events, that conventional prediction methods might misinterpret, and this information can be used to enhance the training dataset.} \deleted{Incorporating feedback from system operators enriches the learning process, especially for predictions marked by high uncertainty. Through iterative cycles of the AL framework, }Each refinement phase aims to \replaced{reduce}{narrow} the forecast error \deleted{margin} and enhance the \deleted{accuracy of }confidence\deleted{intervals}, iteratively improving the model's performance. \deleted{This synergy between the predictive capabilities of the RNN model and the iterative learning process of the AL framework promises to significantly advance power grid management, marking a new era of intelligent, data-informed decision-making.}

\section{Case Study - Results} \label{sec:Results}

This section presents a practical examination of our AI-assisted DT system, enhanced by AL, using the Greek transmission network as a benchmark. We chose this network because it reflects the complexities and challenges that modern power systems face. The case study is designed to demonstrate the effectiveness of our DT \added{architecture} in making accurate \added{and reliable} predictions. \deleted{and improving the management of power grids}.

\subsection{Case Study}

\replaced{Within the scope}{In the realm} of this case study, the Greek transmission network was modeled to establish the core \deleted{structure} of our DT. Using the PyPSA-Eur tool \cite{horsch2018pypsa}, we were able to extract a detailed topological representation of the network, including buses—indicative of substations and generators—and transmission lines. \deleted{, as depicted in Fig. \ref{fig:GreekTN}. This detailed mapping is pivotal, providing a realistic foundation upon which the DT's simulations and analyses are executed.} \added{The DT, enhanced with active learning, turns into a powerful and flexible tool for system operators to manage the power system more effectively, allowing them to use the interface to run different scenarios and supervise the forecasts.}

\ifdraftmode
\else
\fi

To gather the necessary data for the DT, we used a straightforward approach. We gather near real-time and historical data through API calls from ENTSO-E \cite{hirth2018entso} and the Greek TSO IPTO \cite{ipto}. ENTSO-E gives us a wide view of the European grid, while IPTO provides details specific to the Greek system. We also obtain weather data from OpenWeatherMap \cite{openweathermap}, recognizing its impact on the grid's energy demand and production. \deleted{Together, these data sources offer the detailed and timely information our DT needs to effectively simulate and predict grid behavior.} 

\deleted{With all this data, the DT turns into a powerful and flexible tool for those managing the power system. System operators use the interface to work with the DT, running different scenarios and looking at forecasts. This not only helps them understand the current state of the grid better but also helps make the grid stronger and more able to adapt to changes. The active learning part of the DT keeps improving the accuracy of these forecasts by learning from new data as it comes in.}

\subsection{Results}

In our analysis, we developed separate RNN models for each substation within the Greek transmission network, in addition to a model dedicated to forecasting the total load. It is the latter, focusing on the aggregate load across the network, that we showcase here to illustrate the effectiveness of our PLF approach.

Initially, the RNN model was trained using a comprehensive dataset encompassing historical load data and weather conditions across the Greek transmission network. \added{The historical data used for training and testing spans from 2021-01-01 to 2023-12-31. Specifically, the first two years of data (2021-2022) were used for training the RNN model, while the last year (2023) was reserved for testing.} This training set included data on temperature, wind direction, speed, and precipitation, reflecting significant correlations with load patterns as identified through rigorous time-series analysis. \added{The training process for the RNN model took approximately 1 hour on a laptop with the Intel(R) Core(TM) i7-9850H CPU @ 2.60GHz and 16.0 GB RAM.}

\begin{figure}[h]
    \centering
    \includegraphics[width=\columnwidth]{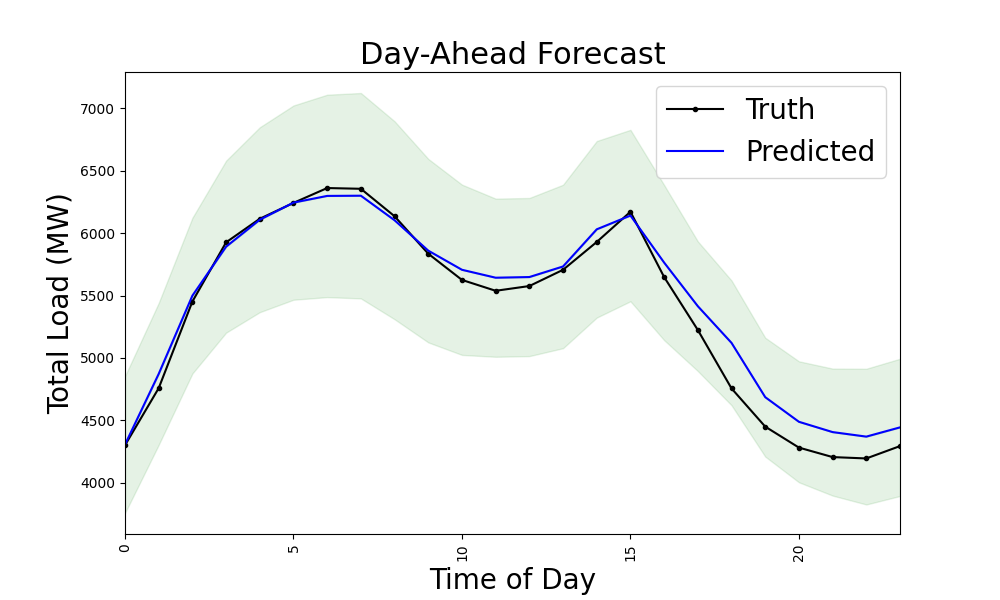}
    \caption{Day-ahead forecast with RNN model together with 95\% confidence intervals.} \label{fig:rnndayahead}
\end{figure}

\deleted{This extensive}\added{The} feature set, coupled with the model's \deleted{sophisticated} architecture—including LSTM layers and dropout for regularization—facilitated a nuanced capture of temporal dependencies.\deleted{and operational dynamics.} \added{The model uses an LSTM as its core network, with one LSTM layer followed by a fully connected layer. The dropout rates are 0.4 for the fully connected layers and 0.3 for the LSTM layer. The activation function used is Leaky ReLU with a leak of 0.1. The training parameters include a maximum of 50 epochs and a batch size of 32.} The model employed a history horizon of 168 hours to inform its predictions, with a forecast horizon set to 24 hours ahead, aligning with the operational requirements for day-ahead planning. Fig. \ref{fig:rnndayahead} provides a visual representation of the model's day-ahead forecasting capabilities, showcasing the precision with which the RNN model can predict total network load alongside the associated 95\% confidence intervals.

We benchmarked the RNN model against traditional forecasting methods such as ARIMA, SARIMA \cite{alberg2018short}, and Prophet \cite{taylor2018forecasting} \added{over the entire test set} to underline its comparative superiority. This qualitative and quantitative comparison highlights the RNN model's enhanced accuracy, reliability, and the ability to produce actionable forecasts, as evidenced by its performance metrics including lower Mean Squared Error (MSE), Root Mean Squared Error (RMSE), and Mean Absolute Error (MAE), alongside improved sharpness and PICP, as can be seen in Table \ref{tab:results}\added{, where the values represent the mean of the metrics over the entire test set.}

\begin{table}[h]
\caption{Performance comparison of forecasting models.}
\label{tab:results}
\resizebox{\columnwidth}{!}{%
\begin{tabular}{|l|l|l|l|l|l|}
\hline
        & MSE    & RMSE   & MAE    & Sharpness & PICP    \\ \hline
ARIMA   & 0.0234 & 0.153  & 0.1282 & N/A    & N/A     \\ \hline
SARIMA  & 0.0107 & 0.1034 & 0.0862 & N/A     & N/A   \\ \hline
Prophet & 0.024  & 0.155  & 0.1357 & 0.3469    & 80.1666 \\ \hline
RNN    & 0.002  & 0.0452 & 0.034  & 0.208     & 97.6774 \\ \hline
\end{tabular}%
}
\end{table}

Notably, the traditional forecasting methods such as ARIMA and SARIMA do not typically provide probabilistic outputs, which is why 'N/A' (not applicable) is listed under the sharpness and PICP columns for these models in Table \ref{tab:results}. This highlights their limitations in providing uncertainty estimates, which are crucial for \replaced{developing the system operator's trust in AI}{efficient power grid management}.

The comparative analysis of forecasting models demonstrates the RNN's superior performance over traditional methods like ARIMA, SARIMA, and Prophet across several key metrics. Qualitatively, the RNN model exhibits significantly higher accuracy and reliability in predicting day-ahead loads, underpinned by its effectiveness in capturing complex temporal dependencies within the data. Furthermore, the RNN model ensures a high level of forecast confidence, as indicated by its competitive sharpness and notably high PICP. This suggests that the RNN model not only predicts with greater accuracy but also provides forecasts with reliable uncertainty estimates, making it a more dependable choice for grid management and operational planning. The integration of an AL framework with the RNN model is anticipated to enhance these attributes further, leveraging real-time data and operator insights for continuous improvement in forecasting performance.

\deleted{The AL process began with the system operator reviewing day-ahead forecasts to identify periods of high uncertainty, particularly those likely to impact grid operations. Using a predefined uncertainty threshold \(\theta\), forecasts exceeding this level were flagged for re-training.  This threshold, empirically set at 1000 based on retrospective analysis, ensures a focused improvement on forecasts with the greatest operational relevance. This methodical selection process, grounded in operational insights, ensured that the AL framework focused on improving forecasts most relevant to grid management challenges.}

\added{The AL process began with the system operator setting the uncertainty threshold \(\theta\) to 1000 based on retrospective analysis, ensuring forecasts with the highest uncertainty were flagged for improvement. The framework then queried actual load values from the data repository, augmented the training set with this data, and retrained the model. After the retraining, the system operator can review and adjust \(\theta\) as needed based on the new confidence intervals. Sensitivity analysis showed that a high \(\theta\) might miss improvement opportunities, since fewer data points will be queried, while a low \(\theta\) could lead to unnecessary computational overhead without significant accuracy gains, since too many data points will be queried.}

\begin{figure}[h]
    \centering
    \includegraphics[width=\columnwidth]{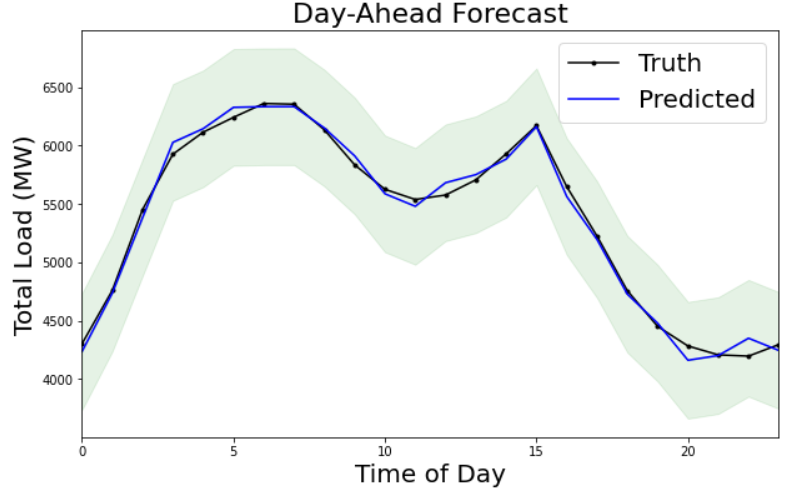}
    \caption{Day-ahead forecast with RNN model together with 95\% confidence intervals after the incorporation of AL.} 
    \label{fig:rnndayaheadal}
\end{figure}

Subsequent re-training of the RNN model with these targeted queries led to observable improvements in forecast accuracy and confidence. For instance, incorporating real-time data corresponding to high-uncertainty predictions enabled the model to adjust to emerging patterns, reducing the overall prediction error and tightening the confidence intervals around forecasts, as can be seen in Fig. \ref{fig:rnndayaheadal}.

To illustrate the enhancements brought about by the AL framework, we present a comparative analysis of the RNN model's performance before and after applying AL. This analysis reveals a marked reduction in forecast error and uncertainty, substantiating the effectiveness of integrating real-time operational feedback into the forecasting process. Table \ref{tab:resultsAL} provides a detailed comparison of key forecasting metrics\added{, presented as the mean of the metrics over the entire test set,} before and after AL enhancement, highlighting the framework's contribution to improved load forecasting accuracy and reliability.

\begin{table}[h]
\caption{Performance comparison of RNN model before and after AL.}
\label{tab:resultsAL}
\resizebox{\columnwidth}{!}{%
\begin{tabular}{|l|l|l|l|l|l|}
\hline
        & MSE    & RMSE   & MAE    & Sharpness & PICP    \\ \hline
RNN (before AL)    & 0.002  & 0.0452 & 0.034  & 0.208     & 97.6774 \\ \hline
RNN (after AL) & 0.001  & 0.0343  & 0.0321 & 0.1823    & 98.3543 \\ \hline
\end{tabular}%
}
\end{table}

After the AL intervention, we observe a decrease in MSE and RMSE, suggesting a tighter fit of the RNN predictions to the true load values. A lower MAE indicates improved average accuracy, and a higher PICP value reflects better coverage of actual loads within the predicted confidence intervals. The decrease in the sharpness value suggests that the confidence intervals have become narrower, indicative of increased precision in the forecasts.

\section{Conclusion and Future Steps} \label{sec:Conlusion}

This study has demonstrated the significant benefits in the accuracy and reliability of day-ahead load forecasts by integrating AI with human system operators through an innovative AL framework and DT architecture. By focusing on the synergy between AI and system operators, we have not only enhanced decision-making processes but also begun to address operator reluctance towards AI adoption. The key to this progress lies in explainability, which builds trust and understanding in AI-generated insights.

Indicative future work includes the expansion of the \replaced{DT architecture's services}{model's capabilities} to encompass broader aspects of power grid optimization, including stochastic, security-constrained, and multi-period \deleted{AC}OPF problems. Furthermore, significant potential is seen in incorporating Large Language Models (LLMs), like ChatGPT for power grid visualization, an example of which is the pioneering ChatGrid platform developed by PNNL \cite{exago}. ChatGrid represents an exciting advancement in generative AI for power grid visualization, offering intuitive, AI-driven insights into grid dynamics, operational constraints, and optimization opportunities. The inclusion of such tools in the presented DT architecture could revolutionize the visualization and interpretation of complex grid data, providing operators with unprecedented clarity and foresight in their decision-making processes.

\section*{Acknowledgment}
This work was supported by European Union’s funded Project HUMAINE [grant number 101120218].



%

\bibliographystyle{unsrt}
\bibliography{references}

\begin{thebibliography}{10}

\bibitem{chen2024artificial}
Yousu Chen, Xiaoyuan Fan, Renke Huang, Qiuhua Huang, Ang Li, and Kishan~Prudhvi Guddanti.
\newblock Artificial intelligence/machine learning technology in power system applications.
\newblock Technical report, Pacific Northwest National Laboratory (PNNL), Richland, WA (United States), 2024.

\bibitem{marot:hal-03123117}
Antoine Marot, Alexandre Rozier, Matthieu Dussartre, Laure Crochepierre, and Benjamin Donnot.
\newblock {Towards an AI assistant for human grid operators}.
\newblock In {\em {Hybrid Human Artificial Intelligence (HHAI)}}, Amsterdam, France, June 2022.

\bibitem{mazumder2024power}
Sudip~K Mazumder, Mohammad Shadmand, H~Alan Mantooth, Chris Farnell, Salam Baniahmed, Arif~I Sarwat, Mohd Tariq, Manimaran Govindrasu, Jay Johnson, and Gab-Su Seo.
\newblock Power grid resilience.
\newblock In {\em Power Electronics Handbook}, pages 1015--1033. Elsevier, 2024.

\bibitem{deakin2023smart}
Matthew Deakin, Marta Vanin, Zhong Fan, and Dirk Van~Hertem.
\newblock Smart energy network digital twins: Findings from a {UK}-based demonstrator project.
\newblock {\em arXiv preprint arXiv:2311.11997}, 2023.

\bibitem{zhou2019digital}
Mike Zhou, Jianfeng Yan, and Donghao Feng.
\newblock Digital twin framework and its application to power grid online analysis.
\newblock {\em CSEE Journal of Power and Energy Systems}, 5(3):391--398, 2019.

\bibitem{gurses2020probabilistic}
Gonca G{\"u}rses-Tran, Hendrik Flamme, and Antonello Monti.
\newblock Probabilistic load forecasting for day-ahead congestion mitigation.
\newblock In {\em 2020 International Conference on Probabilistic Methods Applied to Power Systems (PMAPS)}, pages 1--6. IEEE, 2020.

\bibitem{browell2020probcast}
Jethro Browell and Ciaran Gilbert.
\newblock Probcast: Open-source production, evaluation and visualisation of probabilistic forecasts.
\newblock In {\em 2020 International Conference on Probabilistic Methods Applied to Power Systems (PMAPS)}, pages 1--6. IEEE, 2020.

\bibitem{wang2019deep}
Zengping Wang, Bing Zhao, Haibo Guo, Lingling Tang, and Yuexing Peng.
\newblock Deep ensemble learning model for short-term load forecasting within active learning framework.
\newblock {\em Energies}, 12(20):3809, 2019.

\bibitem{zhang2021active}
Liang Zhang and Jin Wen.
\newblock Active learning strategy for high fidelity short-term data-driven building energy forecasting.
\newblock {\em Energy and Buildings}, 244:111026, 2021.

\bibitem{dagster}
{Elementl}.
\newblock Dagster.
\newblock Software available at https://github.com/dagster-io/dagster.

\bibitem{timescaledb}
{TimeScale, Inc.}
\newblock Timescale{DB}.
\newblock Software available at https://github.com/timescale/timescaledb.

\bibitem{PyPSA}
T.~Brown, J.~H\"orsch, and D.~Schlachtberger.
\newblock {PyPSA: Python for Power System Analysis}.
\newblock {\em Journal of Open Research Software}, 6(4), 2018.

\bibitem{operatorfabric}
{RTE France}.
\newblock Operator{F}abric.
\newblock Software available at https://github.com/opfab/operatorfabric-core.

\bibitem{sherstinsky2020fundamentals}
Alex Sherstinsky.
\newblock Fundamentals of recurrent neural network {(RNN)} and long short-term memory {(LSTM)} network.
\newblock {\em Physica D: Nonlinear Phenomena}, 404:132306, 2020.

\bibitem{paszke2019pytorch}
Adam Paszke, Sam Gross, Francisco Massa, Adam Lerer, James Bradbury, Gregory Chanan, Trevor Killeen, Zeming Lin, Natalia Gimelshein, Luca Antiga, et~al.
\newblock Pytorch: An imperative style, high-performance deep learning library.
\newblock {\em Advances in neural information processing systems}, 32, 2019.

\bibitem{tharwat2023survey}
Alaa Tharwat and Wolfram Schenck.
\newblock A survey on active learning: state-of-the-art, practical challenges and research directions.
\newblock {\em Mathematics}, 11(4):820, 2023.

\bibitem{horsch2018pypsa}
Jonas H{\"o}rsch, Fabian Hofmann, David Schlachtberger, and Tom Brown.
\newblock Pypsa-eur: An open optimisation model of the european transmission system.
\newblock {\em Energy strategy reviews}, 22:207--215, 2018.

\bibitem{hirth2018entso}
Lion Hirth, Jonathan M{\"u}hlenpfordt, and Marisa Bulkeley.
\newblock The {ENTSO-E} transparency platform--a review of {E}urope’s most ambitious electricity data platform.
\newblock {\em Applied energy}, 225:1054--1067, 2018.

\bibitem{ipto}
{IPTO}.
\newblock {IPTO API}.
\newblock \url{https://www.admie.gr/en/market/market-statistics/file-download-api}.

\bibitem{openweathermap}
{OpenWeather}.
\newblock {OpenWeatherMap API}.
\newblock \url{https://openweathermap.org/api}.

\bibitem{alberg2018short}
Dima Alberg and Mark Last.
\newblock Short-term load forecasting in smart meters with sliding window-based arima algorithms.
\newblock {\em Vietnam Journal of Computer Science}, 5:241--249, 2018.

\bibitem{taylor2018forecasting}
Sean~J Taylor and Benjamin Letham.
\newblock Forecasting at scale.
\newblock {\em The American Statistician}, 72(1):37--45, 2018.

\bibitem{exago}
{PNNL}.
\newblock {ExaGO}.
\newblock Software available at https://github.com/pnnl/ExaGO.

\end{thebibliography}

\end{document}